\documentclass[]{aastex631}

\usepackage{multirow}
\usepackage{amsmath}

\newcommand{\swift}{\textit{Swift}}
\newcommand{\tAA}{\text{\AA}}

\begin{document}

\title{EP241217a: a likely Type II GRB with an achromatic bump at $z=4.59$}

\correspondingauthor{Hao Zhou,Zhi-Ping Jin,Da-Ming Wei}
\email{haozhou@pmo.ac.cn,jin@pmo.ac.cn,dmwei@pmo.ac.cn}

\author[0000-0003-2915-7434]{Hao Zhou}
\affiliation{Purple Mountain Observatory, Chinese Academy of Sciences, Nanjing 210023, China}

\author[0000-0002-9037-8642]{Jia Ren}
\affiliation{Purple Mountain Observatory, Chinese Academy of Sciences, Nanjing 210023, China}

\author[0009-0008-8053-2985]{Chen-Wei Wang}
\affil{State Key Laboratory of Particle Astrophysics, Institute of High Energy Physics, Chinese Academy of Sciences, 19B Yuquan Road, Beijing 100049, China}
\affil{University of Chinese Academy of Sciences, Chinese Academy of Sciences, Beijing 100049, China}

\author[0000-0002-4072-6899]{Xing Liu}
\affiliation{National Astronomical Observatories, Chinese Academy of Sciences, Beijing 100101, China}
\affiliation{School of Astronomy and Space Science, University of Chinese Academy of Sciences, Chinese Academy of Sciences, Beijing 100049,People’s Republic of China}

\author[0000-0002-0561-7937]{Bin-Yang Liu}
\affiliation{Purple Mountain Observatory, Chinese Academy of Sciences, Nanjing 210023, China}

\author[0000-0001-7821-9369]{Andrew J. Levan}
\affiliation{Department of Astrophysics/IMAPP, Radboud University Nijmegen, P.O. Box 9010, Nijmegen, 6500 GL, The Netherlands}
\affiliation{Department of Physics, University of Warwick, Coventry, CV4 7AL, UK}

\author[0000-0002-9267-6213]{Jillian Rastinejad}
\affiliation{Center for Interdisciplinary Exploration and Research in Astrophysics (CIERA) and Department of Physics and Astronomy, Northwestern University, Evanston, IL 60208, USA}

\author[0000-0001-9648-7295]{Jin-Jun Geng}
\affiliation{Purple Mountain Observatory, Chinese Academy of Sciences, Nanjing 210023, China}

\author[0000-0002-0556-1857]{Hao Wang}
\affiliation{Purple Mountain Observatory, Chinese Academy of Sciences, Nanjing 210023, China}

\author[0000-0003-0526-2248]{Peter K. Blanchard}
\affiliation{Center for Astrophysics | Harvard \& Smithsonian, 60 Garden St., Cambridge, MA 02138, USA}

\author[0000-0002-7374-935X]{Wen-fai Fong}
\affiliation{Center for Interdisciplinary Exploration and Research in Astrophysics (CIERA) and Department of Physics and Astronomy, Northwestern University, Evanston, IL 60208, USA}

\author[0000-0002-5826-0548]{Benjamin Gompertz}
\affiliation{School of Physics and Astronomy, University of Birmingham, Birmingham B15 2TT, UK}
\affiliation{Institute for Gravitational Wave Astronomy, University of Birmingham, Birmingham B15 2TT}

\author[0000-0002-7517-326X]{Daniele B. Malesani}
\affiliation{Cosmic Dawn Center (DAWN), Denmark}
\affiliation{Niels Bohr Institute, University of Copenhagen, Jagtvej 128, Copenhagen, 2200, Denmark}

\author[0000-0002-5740-7747]{Charles D. Kilpatrick}
\affiliation{Center for Interdisciplinary Exploration and Research in Astrophysics (CIERA) and Department of Physics and Astronomy, Northwestern University, Evanston, IL 60208, USA}

\author[0000-0001-5169-4143]{Gavin P. Lamb}
\affiliation{Astrophysics Research Institute, Liverpool John Moores University, IC2 Liverpool Science Park, 146 Brownlow Hill, Liverpool L3 5RF, UK}

\author[0000-0002-4670-7509]{Brian D. Metzger}
\affiliation{Department of Physics and Columbia Astrophysics Laboratory, Columbia University, New York, NY 10027, USA}
\affiliation{Center for Computational Astrophysics, Flatiron Institute, 162 5th Avenue, New York, NY 10010, USA}

\author[0000-0002-2555-3192]{Matt Nicholl}
\affiliation{Astrophysics Research Centre, School of Mathematics and Physics, Queen’s University Belfast, BT7 1NN, UK}

\author[0000-0003-3274-6336]{Nial R. Tanvir}
\affiliation{School of Physics and Astronomy, University of Leicester, University Road, Leicester LE1 7RH, UK}

\author[0000-0002-8385-7848]{Yun Wang}
\affiliation{Purple Mountain Observatory, Chinese Academy of Sciences, Nanjing 210023, China}

\author[0000-0002-2204-6558]{Yu Rong}
\affiliation{Department of Astronomy, University of Science and Technology of China, Hefei, Anhui 230026, China}
\affiliation{School of Astronomy and Space Sciences, University of Science and Technology of China, Hefei 230026, China}

\author[0000-0001-6223-840X]{Run-Duo Liang}
\affiliation{National Astronomical Observatories, Chinese Academy of Sciences, Beijing 100101, China}

\author[0009-0008-7068-0693]{Zhi-Xing Ling}
\affiliation{National Astronomical Observatories, Chinese Academy of Sciences, Beijing 100101, China}
\affiliation{School of Astronomy and Space Science, University of Chinese Academy of Sciences, Chinese Academy of Sciences, Beijing 100049,People’s Republic of China}

\author[0000-0003-3257-9435]{Dong Xu}
\affiliation{National Astronomical Observatories, Chinese Academy of Sciences, Beijing 100101, China}
\affiliation{Altay Astronomical Observatory, Altay, Xinjiang 836500, People's Republic of China}

\author[0000-0003-4977-9724]{Zhi-Ping Jin}
\affiliation{Purple Mountain Observatory, Chinese Academy of Sciences, Nanjing 210023, China}
\affiliation{School of Astronomy and Space Sciences, University of Science and Technology of China, Hefei 230026, China}

\author[0000-0002-9758-5476]{Da-Ming Wei}
\affiliation{Purple Mountain Observatory, Chinese Academy of Sciences, Nanjing 210023, China}
\affiliation{School of Astronomy and Space Sciences, University of Science and Technology of China, Hefei 230026, China}

\begin{abstract}
EP241217a is an X-ray transient detected by the Einstein Probe (EP) lasting for about 100 seconds and without accompanying $\gamma$-ray detection.
The optical spectroscopy reveals the redshift of EP241217a is 4.59.
By combining the $\gamma$-ray upper limit provided by GECAM-C, there is a considerable possibility that EP241217a is a typical Type II gamma-ray burst (GRB), but it is fainter than the detection threshold of any available $\gamma$-ray monitors (i.e., $E_{\gamma,{\rm iso}}\lesssim10^{53}$\,erg).
The X-ray light curve exhibits a plateau lasting for $\sim5\times10^4$ seconds. However, the joint analysis with optical data suggests the presence of an achromatic bump peaking at $\sim3\times10^4$\,s after the trigger, indicating the actual duration of the X-ray plateau may be significantly shorter than it appears.
To interpret the achromatic bump, we adopt the scenario of a mildly relativistic jet coasting in a wind-like medium and encountering a rapid density enhancement of the circumburst medium, which is likely induced by the the interaction of the progenitor's stellar wind and the interstellar medium.
However, this model cannot fully explain observed data, and some issues do exist, e.g., the observed spectrum is harder than the model prediction.
Consequently, we conclude that the scenario of a mildly relativistic jet coasting in the wind-like medium cannot explain all observed features of EP241217a.
In addition, some alternative models commonly invoked to explain X-ray plateaus are discussed, but there are more or less issues when they are applied to EP241217a.
Therefore, further theoretical modeling is encouraged to explore the origin of EP241217a.

\end{abstract}

\keywords{X-ray bursts (1814)}

\section{Introduction} \label{sec:intro}
The physical origin of fast X-ray transients remains uncertain.
Over the past few decades, analyses based on fluence ratios have suggested the existence of subclasses of gamma-ray bursts (GRBs), namely X-ray-rich GRBs (XRRs) and X-ray flashes (XRFs), which may contribute a significant fraction to the fast X-ray transient population \citep{2001A&A...378..441F,2005ApJ...629..311S,2008ApJ...679..570S,2018ApJ...866...97B}.
XRRs and XRFs have predominantly been identified in gamma-ray triggered events and are found to follow several empirical relations established for classical GRBs  \citep{2018ApJ...866...97B}.
These similarities support the hypothesis that XRRs and XRFs share a common origin with GRBs, albeit involving lower-energy outflows \citep{2005A&A...440..809B,2006ApJ...643..284B,2008MNRAS.388..603L}.
Several theoretical models have been proposed to explain the nature of XRRs and XRFs, including mildly relativistic jets \citep[e.g., the dirty fireball, ][]{2006Natur.442.1014S,2007ApJ...661..982S}, GRBs viewed off-axis \citep{2006ApJ...643..284B,2007ApJ...661..982S,2009A&A...499..439G,2015ApJ...806..222U,2024A&A...690A.101W,2025arXiv250505444G,2025arXiv250304306J}, the spindown-powered emission from a long-lived neutron star \citep{2016ApJ...829...72C} and so on.
Throughout this work, GRBs, XRRs, and XRFs will be generally referred as GRB-like events, which are powered by (likely) catastrophic astrophysical events and involve (mildly) relativistic jets.

Launched in the Jan 2024, the Einstein Probe \citep[EP, ][]{2022hxga.book...86Y} satellite makes it possible to discover fast X-ray transients efficiently with the Wide Field Telescope (WXT, down to a 5-$\sigma$ flux limit of $\sim2.6\times10^{-11}\,{\rm erg\,s^{-1}\,cm^{-2}}$ with exposure time of $1,000$ seconds in the 0.5-4\,keV band).
Once the WXT is triggered by an X-ray transient, EP is capable of performing follow-up observations within a few minutes using the onboard Follow-up X-ray Telescope (FXT) to monitor the temporal and the spectral evolution of the transient.
The Onboard localization uncertainty (90\%) is $\sim3\arcmin$ for WXT and $\sim20\arcsec$ for FXT.
Typically, the onboard WXT(FXT) localization will be transferred to ground within a few seconds(minutes).
After the ground process, the 90\% uncertainty of the FXT localization can be further reduced to $\sim10\arcsec$.
So far, the rate of extra-galactic fast X-ray transients detected by EP is about 2 per week.
A subset of fast X-ray transients has been found to be related with GRB-like events, e.g., EP240315A/GRB 240315C \citep{2025NatAs...9..564L,2025ApJ...979L..28R}, EP240219a/GRB 240219A \citep{2024ApJ...975L..27Y}, EP240801a/XRF 240801B \citep{2025arXiv250304306J} and EP250404a/GRB 250404A \citep{2025arXiv250600435Y}.
Other fast X-ray transients have been linked to core-collapse supernovae, such as EP240414a/SN 2024gsa \citep{2024arXiv241002315S,2025ApJ...978L..21S,2025ApJ...982L..47V,2025ApJ...981...48B} and EP250108a/SN 2025kg \citep{2025arXiv250408886E,2025arXiv250408889R,2025arXiv250417034L,2025arXiv250417516S}, suggesting that these transients may share a common origin with GRBs and some supernovae.
However, a significant number of fast X-ray transients remain of unknown origin. Notable examples include EP240408a \citep{2025SCPMA..6819511Z,2025ApJ...979L..30O}, EP241021a \citep{2025arXiv250505444G,2025arXiv250508781Y,2025arXiv250507665X} and many more waiting for investigation. 

EP241217a is a fast X-ray transient at redshift $z=4.59$, discovered by EP-WXT and not detected by any available $\gamma$-ray instrument.
In this article, we model optical and X-ray data of EP241217a within the framework of the standard fireball-shock model \citep{1998ApJ...497L..17S,1999PhR...314..575P,2003ApJ...597..459Y}.
A ``plateau" is phenomenologically defined as a period when the temporal decay is slower than the behavior inferred from the observed spectrum in the literature \citep{2008MNRAS.383.1143P}. EP241217a exhibits such a plateau lasting for $\sim 5\times10^4$\,s in the X-ray band.
However, when combined optical data, the X-ray plateau lasting for $5\times10^4$\,s is likely an artifact of limited temporal sampling.
Instead, the data indicates the presence of a likely achromatic bump peaking at $\sim3 \times 10^4$ seconds after the trigger, and the end time of the X-ray plateau might be significantly overestimated due to the presence of the achromatic bump.
Moreover, EP241217a appears to be an outlier of the X-ray Luminosity-Time-Energy (LTE) relation (at least 2-$\sigma$)\citep[i.e., a tight three-parameter relation involving the luminosity of the X-ray plateau when it ends $L_{\rm X}$, the end time of the plateau $T_a$, and the isotropic $\gamma$-ray energy of the GRB $E_{\gamma,{\rm iso}}$, ][]{2008MNRAS.391L..79D,2012A&A...538A.134X,2016ApJ...825L..20D,2019ApJ...883...97Z,2019ApJS..245....1T} found in GRB X-ray afterglows with plateaus.

The scenario of a mildly relativistic jet coasting in a wind-like medium \citep{2022NatCo..13.5611D} was adopted to model optical and X-ray light curves of EP241217a, which is also adopted to account for the EP250108a \citep{2025arXiv250417034L}.
In this framework, an achromatic bump can arise if the jet encounters a rapid density enhancement of the circumburst medium \citep{2021ApJ...922...22L}, which might be caused by the interaction between the progenitor’s stellar wind and the surrounding interstellar medium \citep{2006ApJ...643.1036P,2024ApJ...976...55P,2025arXiv250521839Z}.
However, this model does not fully account for all observed features of EP241217a, particularly the spectral properties.
We therefore conclude that the scenario of a mildly relativistic jet coasting in a wind-like medium cannot explain all observed features of EP241217a.
Alternative models for the X-ray plateau, e.g., the energy injection \citep{2001ApJ...552L..35Z,2007ApJ...658L..75G}, the long-lasting reverse shock \citep{2007ApJ...665L..93U,2007MNRAS.381..732G} and so on, are discussed as well.
However, all of these models have more or less issues when applied to EP241217a.

This paper is organized as follows. In Section \ref{sec:obs}, multi-band observations and data reduction of EP241217a are presented.
Section \ref{sec:analysis} presents the temporal and spectral analysis of EP241217a.
The prompt X-ray detection and the $\gamma$-ray upper limit imply there is a considerable possibility that EP241217a is a Type II GRB \citep[originating from a collapsing massive star, ][]{2009ApJ...703.1696Z} that falls below the detection threshold of current $\gamma$-ray monitors.
The afterglow modeling and discussions are presented in Section \ref{sec:origin}.
Finally, our conclusions are summarized in Section \ref{sec:con}.
The standard cosmology model with $H_{0}=67.4\,{\rm km\,s^{-1}\,Mpc^{-1}}$, $\Omega_{M}=0.315$ and $\Omega_{\Lambda}=0.685$ \citep{2020A&A...641A...6P} is adopted in this article.
All uncertainties are reported at the 1-$\sigma$ confidence level unless otherwise stated.

\section{Observations}\label{sec:obs}
\subsection{EP}
EP-WXT was triggered by EP241217a at $T_{\rm trig}=$2024-12-17T05:36:03 (UTC).
About 170 seconds later, the EP-FXT began to perform an autonomous follow-up observation for 3,134 seconds (on-source time throughout the article).
An uncatalogued X-ray source locating at R.A., DEC. = 46.9398, 30.9299 deg with an uncertainty of about 20 arcsec (90\%) was found.
From $\sim8$ hours to $\sim80$ hours after the trigger, EP-FXT performed four more target of opportunity (ToO) observations with exposure time of 3,076\,s, 3,085\,s, 5,892\,s and 9,291\,s, respectively, and EP241217a was detected in the first 3 ToO observations.

\begin{figure}
    \centering
    \includegraphics[width=0.5\linewidth]{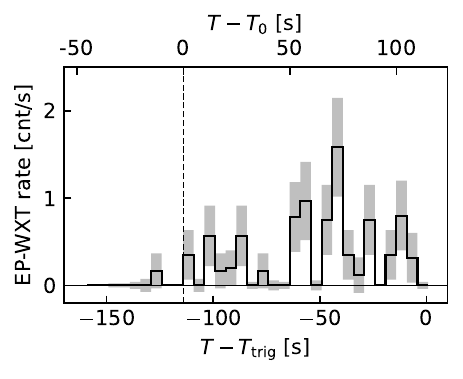}
    \caption{The prompt X-ray light curve of EP241217a. The black line represents the EP-WXT (0.5-4\,keV) rate curve of EP241217a, and the gray region represents the 1-$\sigma$ uncertainty. The width of each time bin is 5\,s. The vertical dashed line represent the possible start time of EP241217a, i.e., $T_0\sim T_{\rm trig}-114.0$\,s.}
    \label{fig:propmt_lc}
\end{figure}

We followed the standard recipe to reduce the WXT data with latest calibration database \citep[CALDB, ][]{2025arXiv250518939C}.
The extracted EP-WXT (0.5-4\,keV) rate curve with a time bin of 5 seconds is shown in Figure \ref{fig:propmt_lc}.
Based on the EP-WXT rate curve, the start time of EP241217a is $T_0\approx T_{\rm trig}-114$\,s.
Pleas note that, this is only a rough estimation.
Because the net count of EP241217a is 47 (distributed over a time range of approximately 114 seconds) and the signal-to-noise ratio (SNR) of the entire period is $\sim6.2$, the WXT light curve may strongly suffer from the background fluctuation (especially for those bins containing only 1 count).
Hence, the first time bin whose 1-$\sigma$ uncertainty does not cover the 0 count/s is selected as the start time.

The EP-FXT data were reduced with \textit{Follow-up X-ray Telescope Data Analysis Software} v1.10 (CALDB v1.10)\footnote{\url{http://epfxt.ihep.ac.cn/analysis}}.
To investigate the spectral evolution of EP241217a, five time-resolved spectra ($<T_0+3.3\times10^4$\,s) were extracted, and the net EP-FXTA count of each spectrum is $\gtrsim200$.
For the temporal analysis, the FXTA/B rate curve is binned with a minimal SNR of 5, but due to the slight difference of sensitivities, edges of time bins are not same for FXTA/B rate curves.
The FXTB rate curve (a bit more sensitive) was resampled to match the time grid of FXTA rate curve, then FXTA/B light curves were combined to one.
As a result, the SNR of each detection points of the combined FXT light curve is $\gtrsim5\sqrt{2}\approx7.1$ per bin, excepting the last 2 detections.
The EP-FXT light curve and spectral indices are shown in Figure \ref{fig:mul_lc}.

The energy conversion factor (ECF, converting the observed rate to the flux or the flux density) was calculated with the best-fit spectral models for epochs with available time-resolved spectra.
For epochs without enough counts to perform robust spectral fitting, the hardness ratio (1.5-10\,keV/0.5-1.5\,keV) is used to calculate the ECF.
The flux or flux density up-limit is calculated with the ECF of the last detection.

\subsection{Swift}
From $\sim T_0+7$ hours to $\sim T_0+22$ hours after the trigger, \textit{Swift} \citep{2004ApJ...611.1005G} observed EP241217a for 3 epochs with exposure time of 1,491.8\,s, 1,752.6\,s and 2,141.2\,s, respectively.
EP241217a was detected by the X-ray telescope (XRT) in the first 2 observations.
The XRT data were reduced with the online XRT product generator \citep{2007A&A...469..379E,2009MNRAS.397.1177E}\footnote{\url{https://www.swift.ac.uk/user_objects}}.
Notably, the first Swift observation is temporally coincident with the first EP-FXT target-of-opportunity epoch, and the derived flux densities at 1,keV from both instruments are consistent within the 1$\sigma$ uncertainties.
The XRT light curve and the spectral indices are shown in Figure \ref{fig:mul_lc}.

\subsection{GECAM-C}
\begin{figure}
    \centering
    \includegraphics[width=0.5\linewidth]{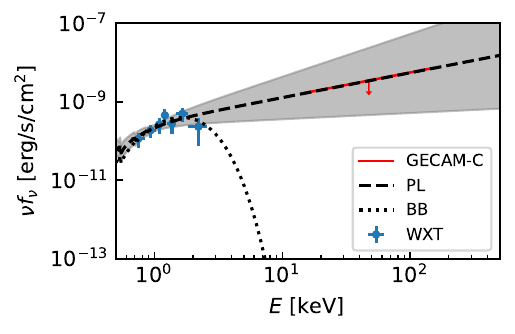}
    \caption{WXT spectrum of the prompt emission. The black dashed line and the black dotted line represent the best fitted PL model and thermal model for the spectrum, and the gray region shows the 1-$\sigma$ uncertainty of the best fitted PL model. The red solid line shows the 3-$\sigma$ upper limit for a 100-second exposure in the 15-150\,keV band, which is almost same as the extrapolation of the best fitted model.}
    \label{fig:wxt_spec}
\end{figure}

Gravitational wave high-energy Electromagnetic Counterpart All-sky Monitor (GECAM) is a constellation with four X-ray and gamma-ray all-sky space telescopes, including GECAM-A/B \citep{GEC_INS_Li2022}, GECAM-C \citep{HEBS_INS_Zhang2023}, and GECAM-D \citep{GTM_INS_wang2024}. 
Among them, the GECAM-C covered the location of EP241217a, and was continuously collecting data throughout the burst duration.
However, no significant $\gamma$-ray signal was detected \citep{GECAM_flt_trigger,GECAM_gnd_search}.
Assuming a power-law model with a photon index of $\Gamma_{\rm X}=1.37$ ($dN/dE\propto E^{-\Gamma_{\rm X}}$, please refer to Section \ref{sec:analysis}), the 3-$\sigma$ up-limit of GECAM-C for a 100-second observation (from $T_0$ to $T_0+100$\,s) is $\sim8.74\times10^{-9}\,{\rm erg/s/cm^2}$ in the 15-150\,keV band, which is shown in Figure \ref{fig:wxt_spec} with the solid red line .

\subsection{Optical observations}
\subsubsection{Gemini-North}
The Gemini-North telescope observed the EP241217a in the $z'$ band at $\sim T_0+2.5$\,h with total exposure time of 600 seconds (PI: Jillian Rastinejad).
The optical counterpart was successfully detected at R.A. = 03:07:46.20 and Dec. = +30:55:45.9 with an uncertainty of $\lesssim0.5^{\prime\prime}$ \citep{2024GCN.38587....1L}.
Follow-up optical spectroscopy was conducted at $\sim T_0 + 4.7$,h with exposure time of 1,600 seconds, confirming the spectroscopic redshift of EP241217a to be $z = 4.59$ \citep[PI: Jillian Rastinejad, ][]{2024GCN.38593....1L}.
As part of the standard observational procedure, several $r'$-band acquisition images were taken prior to spectroscopy.
The $r'$-band acquisition image with the longest exposure time (60\,s) was reduced for photometric analysis, and the optical counterpart of EP241217a is clearly detected.
Aperture photometry was calibrated against nearby Pan-STARRS\,1 stars \citep{2016arXiv161205560C,2020ApJS..251....3M,2020ApJS..251....4W,2020ApJS..251....5M,2020ApJS..251....6M,2020ApJS..251....7F}.

\subsubsection{WFST}
The Wide Field Survey Telescope (WFST, a 2.5\,m survey telescoped located at Lenghu, Qinghai province, China) conducted two $i'$-band follow-up observations at $\sim T_0+32.2$\,h and $\sim T_0+37.1$\,h for 240\,s and 600\,s, respectively.
The optical counterpart of EP241217a was detected in both observations.
WFST data were reduced with the WFST Image Processing Pipeline, an instrument package built on the LSST Science Pipelines v26.0 \citep{Hong_2024,Jenness_2022,Juric_2015}.
Raw images were first processed with Instrument Signature Removal (\texttt{ISR}), which includes bias and overscan subtraction, dark current and flat-field corrections, interpolation over known detector defects, and the generation of a corresponding variance image. 
Each post-ISR exposure then enters \texttt{image characterization}.  
In this stage, the pipeline constructs a spatially varying PSF model (based on the \texttt{PSFEx} algorithm \citep{Bertin_2013}), detects and interpolates cosmic rays, measures an initial sky background, estimates the atmospheric seeing, and refines an initial WCS by matching bright stars to an external reference catalog.  
These calibrations prepare the image for precise astrometry and photometry.
For single-visit calibration, the task \texttt{calibrate} matches well-measured stars to GAIA DR3 \citep{Gaia_2023,Babusiaux_2023,Gaia_2016} and Pan-STARRS\,1 reference catalogs, deriving an accurate SIP+TPV astrometric solution ($\sim$20\,mas\,rms) and an $i$-band photometric zeropoint accurate to 1--2\,\%.  
All flux scaling, throughput curves, and variance factors are stored in a \texttt{PhotoCalib} object attached to the exposure.

Because the fourteen visits overlap on the sky, they are passed to joint calibration (\texttt{jointcal}).
JointCal performs a global astrometric fit across all CCDs and visits and simultaneously refines the relative zeropoints, driving the visit-to-visit photometric scatter down to $\lesssim$10\,mmag and eliminating residual WCS offsets. 
This step places every exposure on a common reference frame and photometric system.
The pipeline then warps each calibrated image onto a common sky tangent-plane grid (tract-patch layout) using a Lanczos-5 kernel.  
A large-scale static sky model is built with 1024-pixel subregion and subtracted from every warp, suppressing broad gradients and CCD-scale discontinuities while preserving low-surface-brightness (LSB) structures.

All background-matched warps are combined in the \texttt{coaddition} step.  
The task \texttt{assembleCoadd} performs an inverse-variance-weighted, sigma-clipped mean stack, flags outlier pixels, and produces a per-pixel variance map.  
To remove any residual background structure, a finer 64-pixel mesh background model was estimated and subtracted from the coadded image.
On the deep-coadded image, the pipeline carries out source detection, deblending, and measurement. 
Objects are detected at signal-to-noise ratios above 5, deblended into child components when necessary, and measured with PSF, CModel, Kron, and fixed-aperture photometry \citep{Bosch_2019}.  
The resulting catalogs are used to perform forced photometry on every original visit, producing light curves for transient candidates such as EP241217a. 

\subsubsection{NOT}
The 2.56-m Nordic Optical Telescope (NOT) observed the field of EP241217a at two epochs.
The first observation was carried out with StanCAM in the $R$ band at $T_0 + 13.64$ h and consisted of 9x180s frames with a median time of $T_0 + 13.91$ h.
The second observation was carried out with ALFOSC in the $i'$ band at $T_0+19.70$d and 9x360s frames were obtained with a median time of $T_0 + 473.24$ h.
The standard image reduction was carried out with \texttt{IRAF} \citep{Tody1986}. 
Astrometric calibration was performed with \texttt{Astrometry.net} \citep{Lang2010} and \texttt{SCAMP} \citep{Bertin2006}.
We performed aperture photometry on the stacked images and calibrate photometries against nearby Pan-STARRS\,1 stars.
The $R$-band photometric zeropoint was calibrated with the photometry converted from the Sloan system \footnote{\url{https://www.sdss4.org/dr12/algorithms/sdssUBVRITransform/\#Lupton2005}}.
The optical counterpart of EP241217a is detected at the first epoch with a magnitude of $R =20.94 \pm 0.05 $mag (Vega), and not detected at the second epoch with a 3-sigma upper limit of $i' = 24.5$ mag (AB).
The magnitudes have not been corrected for galactic dust extinction.

\section{Data analysis}\label{sec:analysis}
The power-law (PL) function is adopted to describe the temporal and spectral behavior of EP241217a, i.e., $f_{\nu}\propto t^{-\alpha}\nu^{-\beta}$, where $\alpha$ and $\beta$ are the temporal decay index and the spectral index, respectively.

\begin{figure}
    \centering
    \includegraphics[width=0.98\linewidth]{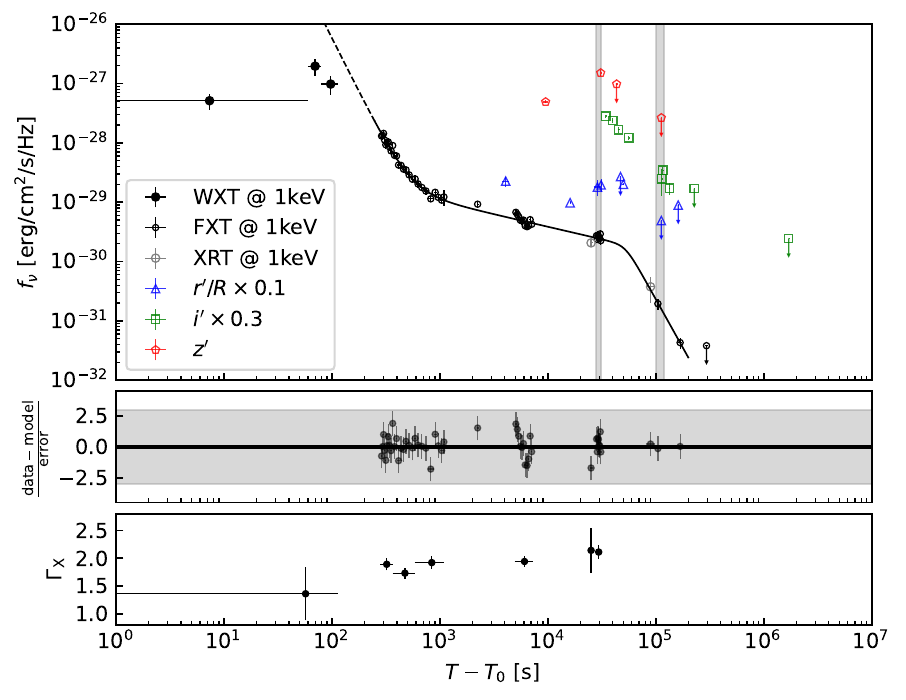}
    \caption{The multi-band light curve of EP241217a. The upper panel shows the optical and X-ray light curves with different markers, and 3-$\sigma$ up-limits are marked with downward arrows. The best fitted phenomenal model, i.e., Equation (\ref{equ:x_model}), for the X-ray data is plotted with the solid black line. The vertical gray regions show the 2 epochs to extract broadband SEDs shown in Figure \ref{fig:sed}. The middle panel shows the residual of the best fitted model for the X-ray data, and the gray region shows the 3-$\sigma$ region. The lower panel shows the X-ray photon index $\Gamma_{\rm X}$, and from $\sim T_0+10^3$\,s to $\sim T_0+10^5$\,s, the photon index is slightly changed from $\sim1.9$ to $\sim2.1$.}
    \label{fig:mul_lc}
\end{figure}

The X-ray light curve of EP241217a can be generally separated into 3 phases: I) $\lesssim T_0+1,000$\,s, the prompt emission ($\lesssim T_0+100$\,s) and the (likely) high-latitude emission \citep[HLE $\gtrsim T_0+100$\,s, ][]{2009MNRAS.399.1328G}, II) $T_0+1,000\,{\rm s}\lesssim T\lesssim T_0+30,000\,{\rm s}$, the shallow decay or the plateau, and III) $\gtrsim T_0+30,000\,{\rm s}$, the steep decay (when $\alpha\gtrsim2$).
Phase I is modeled with a single power-law (PL) decay, while Phases II and III are jointly described using a smoothly broken power-law (SBPL) function, i.e., the temporal decay of the flux density can be modeled by
\begin{equation}\label{equ:x_model}
    f_\nu\propto t^{-\alpha_1}+C\times (\frac{t}{t_{\rm b}})^{-\alpha_2}[1+(\frac{t}{t_{\rm b}})^{1/\Delta}]^{(\alpha_2-\alpha_3)/\Delta},
\end{equation}
where $\alpha_1$, $\alpha_2$, $\alpha_3$ denote temporal decay indices in Phase I, II, and III, respectively, and $t_{\rm b}$ is the break time of the SBPL.
$C$ is a normalization constant to control the intensity ratio of the SBPL component to the PL component, and $\Delta$ governs the smoothness of the break in the SBPL component.
Fitting this model ($\Delta$ fixed to be 0.1, shown as the black line in Figure \ref{fig:mul_lc}) to the X-ray light curve yields best-fit parameters of $\alpha_1=3.62\pm0.25$, $\alpha_2=0.42\pm0.03$, $\alpha_3=3.26\pm0.55$, and $t_{\rm b}=(5.23\pm0.95)\times10^4$\,s. The best-fit model is shown as the black line in Figure \ref{fig:mul_lc}.

The best-fit absorbed PL models to FXT spectra at different epochs indicate that the equivalent hydrogen column density of the host is negligible.
Given the high redshift of EP241217a ($z=4.59)$, the equivalent hydrogen column density of the host ($N_H$) usually cannot be constrained with WXT and FXT data (i.e, the strong absorption feature is below 0.5\,keV), unless $N_H\gtrsim10^{23}\,{\rm cm}^{-2}$.
In addition, the host extinction is $A_V\sim0.25$ (derived from the broadband spectral energy distribution, Section \ref{sec:bump}).
According to the, albeit scattered, observational relation between $N_H$ and $A_V$, the $N_H$ of the host is expected to be $\sim10^{22}\,{\rm cm^{-2}}$ \citep{2013MNRAS.432.1231C,2015A&A...579A..74J}.
Hence, the absorption of the equivalent hydrogen of the host is neglected in our analysis, and the equivalent hydrogen column density of Milky Way is fixed to $1.88\times10^{21}\,{\rm cm^{-2}}$.

\subsection{Prompt emission}
The prompt X-ray spectrum of EP241217a, obtained by WXT, is relatively hard, and we adopted an absorbed PL model and an absorbed blackbody (BB) model to fit the spectrum.
The best-fit spectral index $\Gamma_{\rm X}$ is $1.37^{+0.48}_{-0.45}$ ($dN/dE\propto E^{-\Gamma_{\rm X}}$) for the PL model, and the best-fit temperature $kT$ is $0.4^{+0.09}_{-0.07}$\,keV for the BB model (Table \ref{tbl:sed}).
Although the BB model yields a marginally better fit than the PL model, the improvement in fit quality is not statistically significant \footnote{The likelihood ratio of the BB model to the PL model is about 2.9, which means the two models are almost equivalent.}.
Therefore, we cannot conclude that the BB model is preferred over the PL model based on the current data.

\begin{figure}
    \centering
    \includegraphics[width=0.49\linewidth]{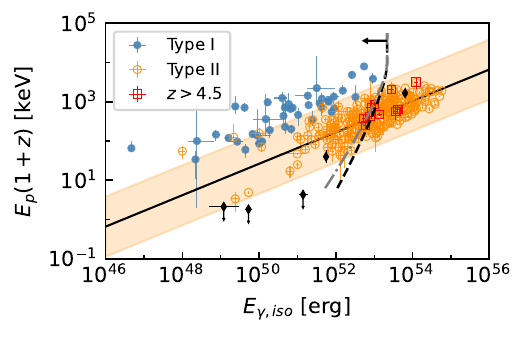}
    \caption{The Amati relation of EP241217a. Black diamonds represent different EP transients, from left to the right are EP250108a, EP240414a, EP241113a, EP240801a and EP240315a, respectively. The GRBs with redshift $>4.5$ are marked with empty red squares, and the black dashed line represents the up-limit of $E_{\gamma, {\rm iso}}$ of EP241217a calculated with the Equation (\ref{equ:x_model}) and Equation (\ref{equ:eiso}). The gray dotted-dashed line represents the Amati relation of EP241217a if assuming a typical high-energy index. The orange region shows the 3-$\sigma$ region of the Amati relation for Type II GRBs. There is a considerable possibility that EP241217a could be a Type II GRB at z=4.59.}
    \label{fig:amati}
\end{figure}

During the prompt phase, EP241217a was covered by the GECAM-C but not detected.
Assuming the best-fitted PL model of the WXT spectrum represents the intrinsic spectrum extending into the $\gamma$-ray regime, the 3-$\sigma$ flux upper limit in the 15-150\,keV band from GECAM-C is $\sim8.74\times10^{-9}\,{\rm erg\,s^{-1}\,cm^{-2}}$ for a 100-second exposure.
The extrapolation of the best-fit PL model of the WXT spectrum and the GECAM-C upper limit is shown in Figure \ref{fig:wxt_spec}.
The GECAM-C upper limit is almost same as the extrapolation of the WXT spectrum, indicating that no stringent constraint can be placed on whether the $\gamma$-ray emission should be detected.
Nevertheless, it is still possible to estimate the upper limit on the isotropic $\gamma$-ray energy $E_{\gamma, {\rm iso}}$ ($1-10^4$\,keV) of EP241217a.
To consider the possible existence of the peak energy $E_p$ within the energy range of $1-10^4$\,keV, the following spectral model is adopted:
\begin{equation}\label{equ:spec_model}
    \frac{dN}{dE}\propto
    \begin{cases}
        E^{-1.37}, & E\le E_p, \\
        E_p^{-1.37}(\frac{E}{E_p})^{-2}, & E>E_p.
    \end{cases}
\end{equation}
The high-energy segment ($E>E_p$) of Equation (\ref{equ:spec_model}) represents a flat $\nu f_\nu$ spectrum, ensuring that $E_{\gamma, {\rm iso}}$ derived with Equation (\ref{equ:spec_model}) will be the largest though it is not physical.
The low-energy segment ($E\le E_p$) of Equation (\ref{equ:spec_model}) represents the best-fit PL model of the WXT spectrum. The $E_{\gamma, {\rm iso}}$ is calculated with
\begin{equation}\label{equ:eiso}
    \begin{aligned}
        E_{\gamma, {\rm iso}}=\frac{4\pi D_L^2}{1+z}T_{\rm d}\int_{E_{\rm lo}/(1+z)}^{E_{\rm hi}/(1+z)}E(\frac{dN}{dE})dE,
    \end{aligned}
\end{equation}
where $D_L$ is the luminosity distance, $T_d$ is the $\gamma$-ray duration of EP241217a (set to 100\,s), $E_{\rm lo}$ and $E_{\rm hi}$ are 1\,keV and $10^4$\,keV, respectively.
The evolution curve of the maximal $E_{\gamma, {\rm iso}}$ with the $E_p$ is shown in Figure \ref{fig:amati} with the gray dashed line.
By comparing this evolution curve with the 3-$\sigma$ region of the Amati relation \citep{2002A&A...390...81A,2020MNRAS.492.1919M} for Type II GRBs and the distribution of Type II GRBs with $z>4.5$ on the Amati diagram, it appears EP241217a has a considerable probability of being a Type II GRB with $E_{\gamma, {\rm iso}}\lesssim10^{53}$\,erg.
This would place it below the detection threshold of currently available $\gamma$-ray monitors.
In addition, if a typical high-energy index of $\sim2.3$ \citep{2021ApJ...913...60P} is set, e.g., $dN/dE\propto E^{-2.3}$ for $E>E_p$, the Amati relation of EP241217a also mainly overlaps with the Type II GRB region (the gray dotted-dashed line in Figure \ref{fig:amati}).
If EP241217a is indeed a Type II GRB, the peak energy in the rest frame can be constrained with the 3-$\sigma$ region of the Amati relation, i.e., $E_p(1+z)\lesssim2.7$\,MeV.
However, if the spectral energy distribution (SED) of EP241217a in the X-ray to $\gamma$-ray band is complex (e.g., X-ray emission is dominated by the thermal component while a distinct component dominates the $\gamma$-ray band), the constraint on the peak energy $E_p(1+z)$ will be incorrect.

\subsection{The late bump}\label{sec:bump}
\begin{figure}
    \centering
    \includegraphics[width=0.49\linewidth]{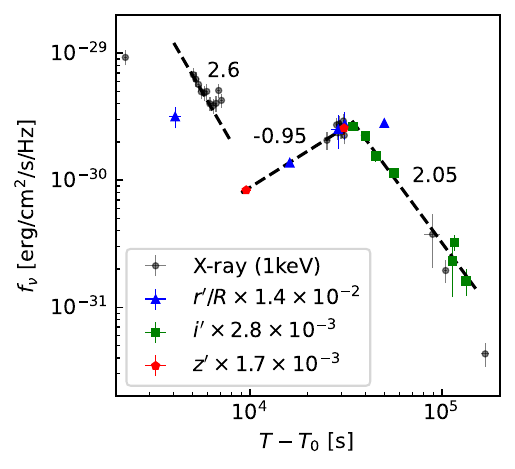}
    \caption{Shifted light curves of EP241217a. All optical light curves are shifted to match the X-ray flux density at $\sim T_0+3\times10^4$\,s. Black dashed lines represent the PL approximation for each segment of shifted light curves, and the numbers represent the temporal decay indices for each approximation. Temporal decay indices of the panchromatic bump at $\sim T_0+3\times10^4$\,s for the rising and the decay phase are $\sim-0.95$ and $\sim2.05$. However, the X-ray light curve decays faster than the optical light curve after the peak, which implies the cooling frequency may locate somewhere between the X-ray and the optical bands. For the X-ray light curve from $\sim T_0+5\times10^3$\,s to $\sim T_0+7\times10^4$\,s, the temporal decay index is about 2.6.}
    \label{fig:merged_lc}
\end{figure}

\begin{figure}
    \centering
    \includegraphics[width=0.5\linewidth]{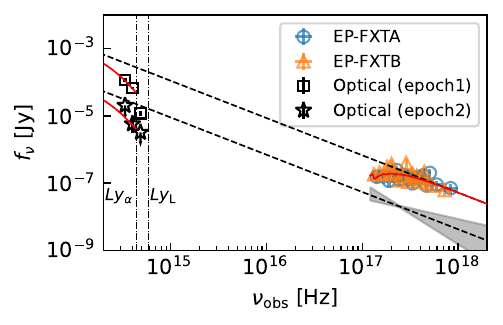}
    \caption{The broadband SEDs of EP241217a. Observed X-ray and optical data are marked with different markers, and black dashed lines represent the best fitted powerlaw model of the X-ray spectrum. For the second epoch, since the count of FXT spectra is not enough to preform reliable spectral analysis, we assumed there is no significant spectral evolution for the 2 epochs. Red solid lines represent the model predicted observed values. The intrinsic extinction is assumed to be similar to SMC, and the V-band host extinction is $A_V^{\rm host}\approx0.25$. The 2 dot-dashed lines represent the Lyman $\alpha$ (1216\,\tAA) and Lyman limit (912\,\tAA).}
    \label{fig:sed}
\end{figure}

During Phase II, the optical light curve of EP241217a is sparsely sampled in the $r'$ and $z'$ bands, while Phase III is well sampled in the $i^\prime$ band.
Both $z^\prime$-band and $r^\prime/R$-band light curves exhibit a bump peaking at $\sim T_0+3\times10^4$\,s.
By shifting optical light curves to match the intensity of the X-ray emission at $\sim T_0+3\times10^4$\,s, the shifted multi-band light curves imply the bump is likely achromatic (shown in Figure \ref{fig:merged_lc}).
However, the epoch of the key Swift-XRT observation ($\sim T_0+2.5\times10^4$\,s), which may indicate the presence of a bump in the X-ray band, is too close to the subsequent EP-FXT observation ($\sim T_0+3.0\times10^4$\,s).
As a result, the observed X-ray flux density (1\,keV) does not exhibit a statistically significant deviation from the best-fit model (the lower panel of Figure \ref{fig:mul_lc}).
Consequently, if the X-ray bump is real, the late bump of EP241217a is achromatic, suggesting a likely external origin, e.g., the rapid enhancement of the density of the circumburst medium (Please see Section \ref{sec:origin} for details).
Temporal decay indices during the rising and the decay phases of the bump are estimated to be $\sim-0.95$ and $\sim2.05$, respectively.
To further investigate the spectral evolution, two epochs with middle time of $\sim T_0+3.0\times10^4$\,s and $\sim T_0+1.1\times10^5$\,s (shown as vertical gray regions in Figure \ref{fig:mul_lc}) were selected to extract broadband spectral energy distributions of EP241217a (Figure \ref{fig:sed}).
For the second epoch, the net counts ($\sim17$ for FXTA and $\sim19$ for FXTB) are not enough to perform reliable spectral fitting.
Therefore, the hardness ratio HR(1.5-10\,keV/0.5-1.5\,keV) of FXT data is calculated to derive the photon index of the second epoch.
The observed HR(1.5-10\,keV/0.5-1.5\,keV) of the second epoch is $0.52\pm0.19$, corresponding to a photon index of $\sim2.0$, which is marginally consistent with the spectral index of the first epoch.
Hence, we assume the X-ray spectrum did not evolves significantly between the two epochs.

Compared with the first epoch, the X-ray flux density (1\,keV) at the second epoch becomes fainter by a factor of $\sim0.08$, which is derived from the best fitted phenomenological model of the X-ray light curve, i.e., Equation (\ref{equ:x_model}).
Assuming the best-fit PL model of FXT data is the intrinsic SED extending to the optical band of EP241217a, the extinction of the host is determined to be $A_V\sim0.25$, under the assumption that the extinction curve of the host is similar to the extinction curve of the Small Magellanic Cloud (SMC).
The 2 broadband SEDs (optical to X-ray) of EP241217a show that there is no obvious spectral evolution during the 2 epochs.

\section{Physics of EP241217a}\label{sec:origin}
We assume that EP241217a is a GRB-like event, i.e., involving a (mildly) relativistic jet interacting with the circumburst medium.
Accordingly, we attempt to interpret observational properties of EP241217a within the framework of the standard fireball-shock model \citep{1998ApJ...497L..17S,1999PhR...314..575P,2003ApJ...597..459Y}.
There are already several well developed algorithms to calculate the afterglow emission numerically, like \texttt{afterglowpy} \citep{2020ApJ...896..166R,2024ApJ...975..131R}, \texttt{jetsimpy} \citep{2024ApJS..273...17W}, \texttt{ASGARD} \citep{2024ApJ...962..115R}, \texttt{VegasAfterglow} \citep{2025arXiv250710829W}\footnote{\url{https://github.com/YihanWangAstro/VegasAfterglow}} and a library of public codes \texttt{redback} \citep{2024MNRAS.531.1203S}.

The correction factor converting the observed $r^\prime$-band flux densities to the unabsorbed values is $\sim19.7$, which is derived from the broadband SED at the first epoch (Figure \ref{fig:sed}).
The unabsorbed $r^\prime$-band flux density at $\sim T_0+4,000$\,s is thus estimated to be $f_\nu(r^\prime)\sim2.9\times10^{-27}\,{\rm erg\,s^{-1}\,cm^{-2}\,Hz^{-1}}$.
Extrapolating the X-ray light curve (assuming the temporal decay index is $\sim2.6$, as shown in Figure \ref{fig:merged_lc}) to $\sim T_0+4\,000$\,s yields an estimated X-ray flux density $f_\nu(1\,{\rm keV})\sim1.18\times10^{-29}\,{\rm erg\,s^{-1}\,cm^{-2}\,Hz^{-1}}$.
Hence, the optical to X-ray spectral index at $\sim T_0+4\,000$\,s is $\beta_{\rm OX}\sim0.88$, corresponding to $\Gamma_{\rm OX}\sim1.88$, which is consist with $\Gamma_{\rm X}=1.94\pm0.09$ determined from the subsequent FXT spectrum.
From $\sim T_0+4\times10^3$\,s, optical and X-ray emissions may have the same origin.

If the progenitor of EP241217a is a collapsing massive star, the interaction of the stellar wind and the interstellar medium (ISM) is expected to give rise to a shock structure, which can naturally produce a rapid density enhancement \citep{2006ApJ...643.1036P,2024ApJ...976...55P,2025arXiv250521839Z}.
From $\sim T_0+700$\,s to $\sim T_0+3\times10^4$\,s, the X-ray spectral index $\beta_{\rm X}$ evolves from $\sim0.9$ to $\sim1.1$.
The reason for the minor change of the spectral index is not clear.
One possible explanation is the passage of the cooling frequency $\nu_{\rm c}$ \citep[a very smooth break, ][]{2014ApJ...780...82U}.

\subsection{Afterglow modeling}In the standard fireball framework, the bump at $\sim T_0+3\times10^4$\,s might be the onset of the afterglow.
After the peak, the temporal decay index of the i-band light curve is $\alpha\sim2$ (Figure \ref{fig:merged_lc}) and the X-ray spectral index is $\beta_{\rm X}\sim1.1$ (Table \ref{tbl:fit_par}).
For the spectral scenario $\nu_{\rm m}<\nu_{\rm X}<\nu_{\rm c}$, the electron index derived from the X-ray spectral index is $p_\beta=1+2\beta_{\rm X}\approx3.2$.
The electron index can be estimated with the temporal decay index as well, which is denoted as $p_\alpha$.
Before the jet break, the electron index can be estimated with $p_\alpha=1+4\alpha/3\approx3.7$ and after the jet break $p_\alpha=4\alpha/3\approx2.7$.
Hence, for the spectral scenario $\nu_{\rm m}<\nu_{\rm X}<\nu_{\rm c}$, the spectrum and the light curve can not derive a proper electron index.
For the spectral scenario $\nu_{\rm X}>\nu_{\rm c}$, $p_\beta=2\beta_{\rm X}\approx2.2$.
Before and after the jet break, $p_\alpha=(4\alpha+2)/3\approx3.3$ and $p_\alpha=(4\alpha-1)/3\approx2.3$, respectively.
Hence, after $\sim T_0+3\times10^4$\,s, the most possible solution is that the cooling frequency had passed through the X-ray band and the jet break had happened in the standard fireball framework, which gives an electron index of $\sim2.2-2.3$.
As a result, the bump around $\sim T_0+3\times10^4$\,s could be a comprehensive result of the onset of the afterglow and the jet break.

The onset of the afterglow can be used to estimate the initial bulk Lorentz factor, and the deceleration time (i.e., when the bulk Lorentz factor is about half of the initial value) approximately equals the peak time of the light curve, i.e., $t_{\rm dec}\approx t_{\rm peak}\sim 3\times10^4\,{\rm s}$.
In the standard fireball framework, the initial bulk Lorentz factor can be estimated with $\Gamma_0\approx39(\frac{t_{\rm dec}}{3\times10^4\,{\rm s}})^{-3/8}(\frac{1+z}{1+4.59})^{3/8}(\frac{E_{\gamma,{\rm iso}}}{10^{53}\,{\rm erg}})^{1/8}(\frac{\eta_\gamma}{0.2})^{-1/8}(\frac{n}{1\,{\rm cm^{-3}}})^{-1/8}$ for the ISM scenario, $E_{\gamma,{\rm iso}}$ is the isotropic $\gamma$-ray energy of the jet, $\eta_\gamma$ is the factor that the kinetic energy converted into $\gamma$ photons and $n$ is the density of the ISM.
In addition, as discussed above, the jet break time is also approximately equal to the peak time, i.e., $t_{\rm jet}\approx t_{\rm peak}\sim3\times10^4\,{\rm s}.$
Thus, the half jet opening angle $\theta_j\sim2.0^\circ(\frac{t_{\rm peak}}{3\times10^4\,{\rm s}})^{3/8}(\frac{1+z}{1+4.59})^{-3/8}(\frac{E_{\gamma,{\rm iso}}}{10^{53}\,{\rm erg}})^{-1/8}(\frac{\eta_\gamma}{0.2})^{1/8}(\frac{n}{1\,{\rm cm^{-3}}})^{1/8}$\citep{2001ApJ...562L..55F}.

As a result, the standard fireball framework predicts the jet of EP241217a may be mildly relativistic with an initial Lorentz factor of a dozens and the half jet opening angle of a few degrees.
The afterglow model was fitted to the data after $T_0+1,000$\,s, and for X-ray data before $T_0+1,000$\,s, a PL decay with $\alpha=3.63$ was adopted.
Firstly, we tried to model observed light curves in the standard framework \citep{1998ApJ...497L..17S} for the ISM scenario, but it failed to simultaneously explain the shallow decay phase in the X-ray band from $\sim10^3$\,s to $\sim10^4$\,s and the bump at $\sim T_0+30\,000$\,s.
\citep{2020ApJ...893...88O} proposed that the shallow decay might be induced by the high-latitude emission (HLE) of the structured jet.
The HLE usually dominates in the X-ray band and is much fainter than the forward shock in the optical band.
However, for EP241217a, the HLE should be much brighter than the forward shock in the optical band to account for the first R/r-band detection and comparable with the X-ray emission, which conflicts with Oganesyan’s prediction.

For the wind scenario, the transition from the coasting to the deceleration is more likely to produce a transition from the shallow decay to the normal decay instead of a bump, which is discussed in Appendix \ref{app:jet}.

Hence, the following scenario was assumed to account for EP241217a.
The progenitor is a collapsing massive star, and the interaction between the stellar wind and the ISM leads to a shock structure, which naturally produces a rapid density enhancement of the circumburst medium.
A relativistic jet is launched after the collapse of the massive star, and interacted with the circumburst medium.
The X-ray plateau is interpreted as the emission of the jet coasting in a wind-like medium, and the achromatic bump is attributed to the jet encountering the density enhancement of the circumburst medium.
Finally the jet enters the shocked wind medium \citep[ISM-like,][]{2024ApJ...976...55P,2025arXiv250521839Z}.
A jet break occurs after the transition to the ISM-like medium.
Consequently, for $T\gtrsim T_0+3\times10^4$\,s, the temporal decay index is $\alpha\gtrsim2$ for the X-ray and the optical bands.

We employ the \texttt{ASGARD} program as the foundation for fitting the afterglow light curves. The physical processes underlying this program are outlined in the Appendix of \cite{2024ApJ...962..115R}. The following modules within this package are invoked for the present work: external-forward shock dynamics, synchrotron and synchrotron self-Compton (SSC) radiation, electron energy spectrum evolution incorporating inverse Compton (IC) cooling effects, and the equal arrival time surface effect. The physical scenario is configured as a top-hat jet propagating in a free stellar wind transitioning to shocked wind environment, with jet lateral expansion neglected. We utilize the \texttt{PyMultiNest} program as the sampler to perform nested sampling, fitting the theoretical light curves generated by \texttt{ASGARD} to the observational data. Additional corrections for optical extinction, as described in Appendix~\ref{app:lyman}, are applied. 
Finally, the sampling was carried out in a 12 dimensional parameter space.
The model parameters include
\{$n_0$, $A_{\star}$, $E_{\rm k,iso}$, $p$, $\Gamma_{0}$, $\epsilon_{e}$, $\epsilon_{B}$,
$\xi_{e}$, $\theta_{\rm j}$,
$\theta_{\rm v}/\theta_{\rm j}$, $A_r$, $E(B-V)$\}.
The meaning of each parameter is given in Appendix~\ref{app:par}.

\begin{figure}
    \centering
    \includegraphics[width=0.98\linewidth]{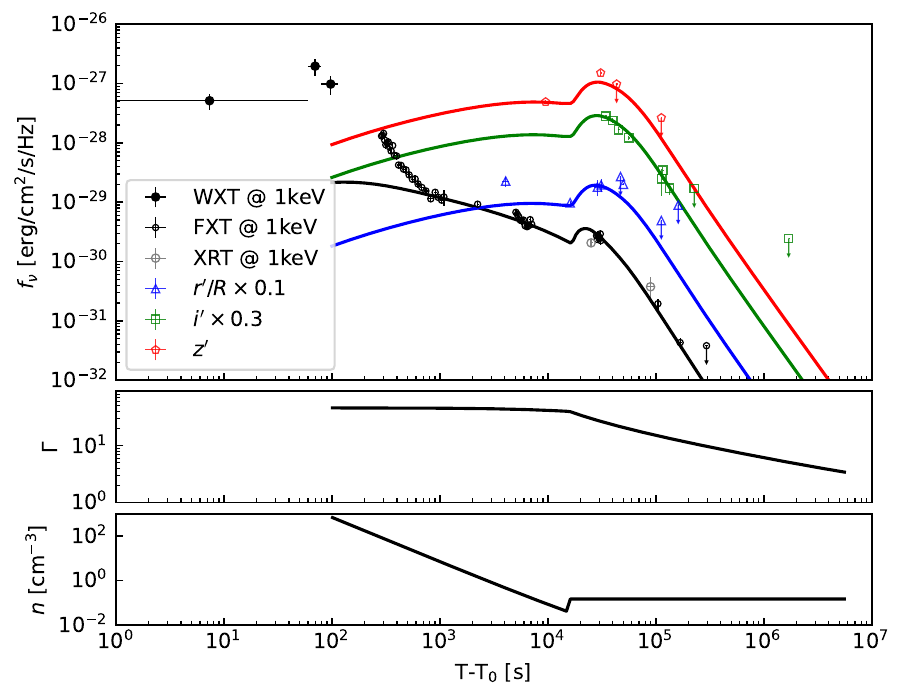}
    \caption{The best-fit afterglow model of EP241217a. The upper panel shows the best-fit multi-band light curves of EP241217a. The middle panel shows the bulk Lorentz factor of the jet. The lower panel shows the density profile of the circumburst medium.}
    \label{fig:modelLc}
\end{figure}

The fitted light curves, the bulk Lorentz factor of the jet and the density profile of the circumburst medium are plotted in Figure \ref{fig:modelLc}.
The fitting result indicates that the initial Lorentz factor of the jet is $\Gamma_0=46^{+43.4}_{-14.8}$, and EP241217a was viewed slightly off-axis with a half opening angle of $\theta_j\sim0.77^\circ$ and a viewing angle of $\theta_v\sim1.32^\circ$.
Hence, the lack of strong prompt $\gamma$-ray emission is naturally explained.
The isotropic kinetic energy of the jet reaches $E_{\rm k, iso}\sim1.7\times10^{55}$\,erg, thus the efficiency extracting the kinetic energy to $\gamma$-ray photons is extremely low (i.e., $E_{\gamma, {\rm iso}}/E_{\rm k, iso}\lesssim1\%$), which is consistent with the slightly off-axis scenario.
The electron index is constrained to be $p=2.81^{+0.08}_{-0.10}$.
The best fitted parameters are summarized in Table \ref{tbl:fit_par} and the detailed corner plot is in Appendix \ref{app:par}.

Before the jet encountering the rapid density enhancement of the circumburst medium (i.e., $\sim T_0+1.6\times10^4$\,s), the jet was coasting in the wind-like medium (i.e., $\Gamma\propto t^0$).
During the coasting phase, the model predicts an X-ray plateau but the predicted spectrum $\sim\nu^{-1.4}$ is softer than observed $\sim\nu^{-0.9\pm0.1}$.
Detailed discussions are presented in Section \ref{sec:discussion}, and we found that there are some issues to explain observed features of EP241217a in the scenario of a mildly relativistic jet coasting in a wind-like medium.

\subsection{The X-ray plateau}
A plateau is phenomenologically defined when the temporal decay is slower than the prediction derived from the spectrum \citep[typically $\alpha\lesssim0.5$, ][]{2008MNRAS.383.1143P}, and a tight three-parameter relation for X-ray plateaus in GRBs has been found \citep{2008MNRAS.391L..79D,2012A&A...538A.134X,2016ApJ...825L..20D,2019ApJ...883...97Z,2019ApJS..245....1T}, namely, the LTE relation involving the luminosity of the X-ray plateau when it ends $L_{\rm X}(0.3-10\,{\rm keV})$, the end time of the plateau $T_a$ and the isotropic $\gamma$-ray energy of the GRB $E_{\gamma,{\rm iso}}(15-150\,{\rm keV})$ in the rest frame.

According to the phenomenalogical fitting result of the X-ray light curve in Section \ref{sec:analysis}, the plateau of EP241217a lasts for $\sim5.2\times10^4$\,s. The X-ray luminosity at the end of the plateau is calculated with the following equation
\begin{equation}
    L_{\rm X}(0.3-10\,{\rm keV})=4\pi D_L^2\int_{0.3\,{\rm keV}/(1+z)}^{10.0\,{\rm keV}/(1+z)}f_\nu(E)dE,
\end{equation}
where $D_L$ is the luminosity distance, $f_\nu(E)\propto E^{-\beta_{\rm X}}$ is the flux density at the end of the plateau ($\beta_{\rm X}$ is set to 1.1).
For EP241217a, the X-ray rest-frame luminosity is $L_{\rm X}(0.3-10\,{\rm keV})=(3.4\pm1.0)\times10^{47}$\,erg/s.
The end time of the plateau was assumed to be same as the $T_a=t_b/(1+z)$, which is obtained in Section \ref{sec:analysis}, and the 3-$\sigma$ upper limit of the isotropic $\gamma$-ray energy (15-150\,keV) is $E_{\gamma, {\rm iso}}=1.17\times10^{52}$\,erg, which is derived from the GECAM-C observation.

\begin{figure}
    \centering
    \includegraphics[width=0.5\linewidth]{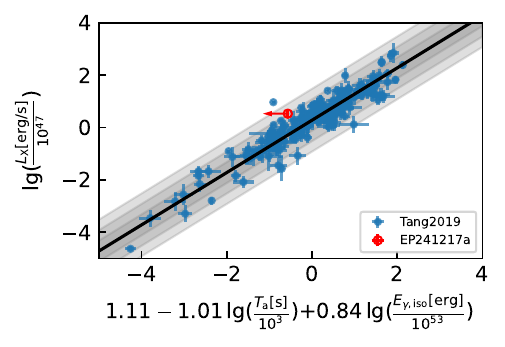}
    \caption{LTE relationship of X-ray plateaus in GRB afterglows. EP241217a is marked with the red empty circle. The blue points are taken from \cite{2019ApJS..245....1T}, and the black solid line with gray shaded regions represents the best fitted model with 1-$\sigma$, 2-$\sigma$ and 3-$\sigma$ uncertainties.}
    \label{fig:xp}
\end{figure}

EP241217a appears to be an outlier of the LTE relation found in GRB afterglows (Figure \ref{fig:xp}) for at least 2-$\sigma$.
However, when combining the optical data, there could be an achromatic bump at $\sim T_0+3\times10^4$\,s in the X-ray band.
The end time of the X-ray plateau may be overestimated due to the achromatic bump.
In addition, it was believed that the jet is viewed on-axis for most GRBs, hence, in principle, it is necessary to correct the observed $E_{\gamma, {\rm iso}}$ of EP241217a to the on-axis scenario.
If the true end time ($T_a$) of the plateau is shorter and/or the true $E_{\gamma, {\rm iso}}$ is larger, the possibility that EP241217a conforms the LTE relation will increase (Figure \ref{fig:xp}).

Since the scenario of a (mildly) jet coasting in a wind-like medium is not sufficient to fit to data, we argue that it is unlikely to fully account for the X-ray plateau observed in EP241217a.
A more detailed discussion of the limitations of this interpretation is presented in the following section.

\subsection{Alternative models}\label{sec:discussion}
\begin{figure}
    \centering
    \includegraphics[width=0.98\linewidth]{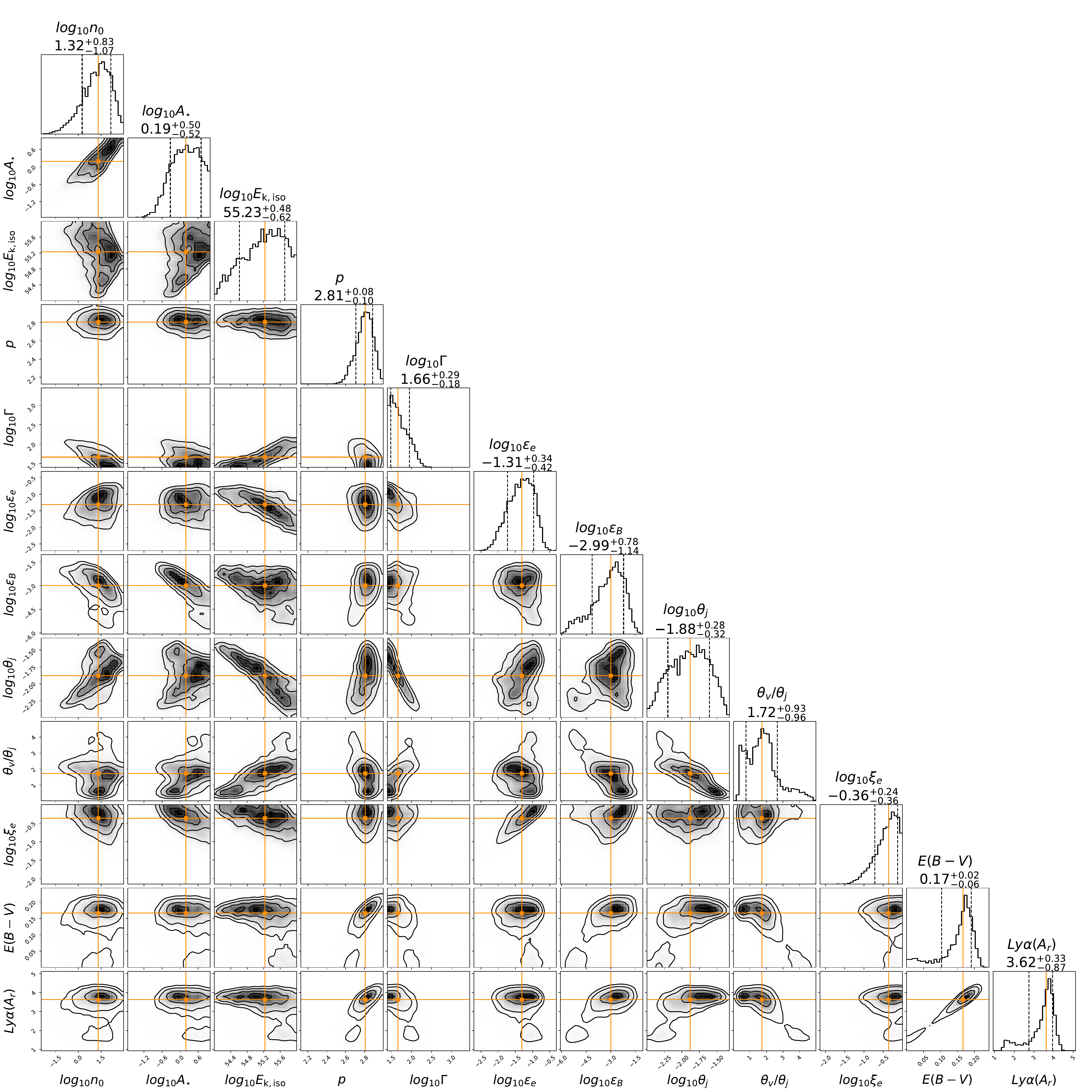}
    \caption{Posterior distributions of fitted parameters. All uncertainties are reported at 1-$\sigma$ confidence level. The meaning of all parameters are described in details in Appendix \ref{app:par}.}
    \label{fig:corner}
\end{figure}

Base on the best-fit parameters, the synchrotron emission during the X-ray plateau is in the fast cooling regime.
In this scenario, the X-ray band lies above both the cooling frequency $\nu_c\sim(6.6\times10^{13}\,{\rm Hz})(t/5,000\,{\rm s})$ and the typical emission frequency $\nu_m\sim(1.8\times10^{15}\,{\rm Hz})(t/5,000\,{\rm s})^{-1}$.
As a result, the model predicts the X-ray flux density should follow the scaling relation $f_\nu(1\,{\rm keV})\propto t^{-(p-2)/2}\nu^{-p/2}$ (please refer to Appendix \ref{app:jet} for details).
With the best-fit electron index $p\sim2.8$, during the X-ray plateau, the scaling relation is $f_\nu(1\,keV)\propto t^{-0.4}\nu^{-1.4}$.
The observed temporal index during the plateau ($\alpha_2=0.42\pm0.03$) is consistent with the prediction, but the observed X-ray spectral index ($\beta_{\rm X}=0.94\pm0.09$) is harder than the prediction.
In addition, the X-ray spectrum exhibits a slight softening over time ($\Gamma_{\rm X}\sim1.9$ to $\sim2.1$), however, the model predicts a constant $\Gamma_{\rm X}$ or a hardening of $\Gamma_{\rm X}$.
In addition, the extremely small jet opening angle \citep{2021ApJ...912...95R,2023ApJ...959...13R} and the high isotropic kinetic energy appear to be in tension with the low initial Lorentz factor \citep{2025ApJ...991..209Z}.
The collimation-corrected energy, given by $E_k=(1-{\rm cos}\theta_j)E_{k, iso}\approx1.5\times10^{51}\,{\rm erg}$, represents the true energy released in the event, and the corresponding prompt energy in the gamma-ray band is $E_\gamma\sim(3\times10^{50}\,{\rm erg})(\frac{\eta_\gamma}{0.2})(\frac{E_k}{1.5\times10^{51}\,{\rm erg}})$, which is consistent with the typical value \citep{2004ApJ...616..331G}.
Given the strong anti-correlation between isotropic kinetic energy and jet opening angle (see Figure \ref{fig:corner}), a larger jet opening angle (i.e., the jet break happened later than the bump) would naturally lead to a lower collimation-corrected energy, thereby resolving the apparent tension.
These inconsistencies indicate that, although the scenario that a mildly relativistic jet coasting in a wind-like medium fails to account for all the observed features of EP241217a.
Therefore, this scenario is unlikely to be the sole explanation for the event.

From Phase II to Phase III, the temporal decay index changes significantly from $\sim0.4$ to $\gtrsim2.1$.
To account for the steep decay during Phase III, the jet break is assumed to happen at $\sim T_0+3\times10^4$\,s.
The best-fit model reveals the jet is coasting up to $\sim T_0+1.6\times10^4$\,s, hence the bulk Lorentz factor around the jet break is estimated to be $\Gamma_j\approx \Gamma_0 (t/t_0)^{-(3-k)/(8-2k)}=46\times(3\times10^4/1.6\times10^4)^{-(3-k)/(8-2k)}$ \citep{2012ApJ...744...36S}, where $k$ is the index of the circumburst density profile (i.e., $n\propto r^{-k}$. For the ISM ($k=0$) medium, $\Gamma_j$ is $\sim36$, and for the wind ($k=2$) medium, $\Gamma_j$ is $\sim39$.
The jet break means the jet opening angle is approximately equals to $1/\Gamma_j$, hence the half opening angle of the jet can be constrained with $\theta_j\sim(180/\pi\Gamma_j)^\circ\approx1.5^\circ$.
 Consequently, a smaller initial bulk Lorentz factor of the jet will lead to a larger jet opening angle.

EP241217a appears to be an outlier of the LTE relation found in GRB X-ray afterglows for at least 2-$\sigma$.
However, our analysis reveals that the end time of the X-ray plateau of EP241217a is overestimated (due to the likely achromatic bump) and the isotropic $\gamma$-ray energy is underestimated (the jet is viewed slightly off-axis).
If applying proper correction to $T_a$ and $E_{\gamma, {\rm iso}}$, the possibility that the X-ray plateau of EP241217a follows the LTE relation will increase considerably.

Excepting the scenario of a mildly relativistic jet coasting in a wind-like medium, there are other models to explain the X-ray plateau found in GRB afterglows, which are mainly divided into three categories:
1) the continuous energy injection from the central engine or the continuous outflow \citep{1998PhRvL..81.4301D,1998A&A...333L..87D,2001ApJ...552L..35Z,2006MNRAS.372L..19F,2007ApJ...658L..75G,2013ApJ...779...28G,2014MNRAS.442.3495V}.
2) a long-lasting reverse shock \citep{2007ApJ...665L..93U,2007MNRAS.381..732G}.
3) a jet viewed off-axis or the HLE emission of a structured jet \citep{2020ApJ...893...88O,2020A&A...641A..61A}.
In the energy injection scenario, the injected luminosity is commonly parameterized as $L_i \propto t^{-q}$.
Assuming the central engine of EP241217a is a magnetar, the injected luminosity follows $L_i\propto(1+t/t_i)^{-2}$, where $t_i$ is the spin down time scale.
The injected luminosity is almost a constant (i.e., $q=0$) during the plateau, and after the plateau, the injected luminosity decays as $t^{-2}$.
Since the broadband spectral index (optical to X-ray) is approximately equal to the X-ray spectral index during Phase II and Phase III, the spectral band from optical to X-ray may belong to the same segmentation of the synchrotron spectrum, e.g., $\nu_m<\nu_{\rm O}<\nu_{\rm X}<\nu_c$.
The electron index can be derived with the X-ray spectral index during Phase II (i.e., $p=2\beta_{\rm X}+1\approx2.8$), if the X-ray frequency satisfies$\nu_{\rm X}<\nu_c$.
The slight softening of the spectrum could be induced by the passage of the cooling frequency \citep[a very smooth break, ][]{2014ApJ...780...82U}.
As a result, the scaling relation (ISM) is expected to be $f_\nu(1\,{\rm keV})\propto t^{-(3(p-1)/4+(p+3)(q-1)/4)}$ for Phase II and $f_\nu(1\,{\rm keV})\propto t^{-((3p+1)/4+(p+2)(q-1)/4)}$ for Phase III (after the jet break), respectively.
To produce observed decays, the injection index is estimated to be $q\approx0.35(1.8)$ during Phase II(III), which falls in the tolerant range of the magnetar hypothesis (i.e., $0<q<2$).
Nevertheless, to produce the achromatic bump, a rapid enhancement of the density of the circumburst medium may be required, in addition to energy injection, which makes the model a bit over tuned.
In the long-lasting reverse shock hypothesis, the chromatic behavior is expected \citep{2007ApJ...665L..93U}, however, the temporal behavior of EP241217a seems to be achromatic.
Furthermore, if the achromatic bump and the following steep decay are dominated by the reverse shock, the emission from the forward shock is expected during Phase III , which decays slower than the reverse shock emission but is not observed.
Regarding the structured jet viewed slightly off-axis or on-axis hypothesis, \cite{2020ApJ...893...88O} demonstrated that the late-time HLE emission of a Gaussian jet can produce a X-ray bump at $\sim10^3$\,s, but the optical bump should not be observed, which is conflict with observations.
Therefore, for EP241217a, current models explaining X-ray plateaus, while successful in some cases, have more or less issues when applied to account for EP241217a.

A refreshed shock is also able to account for the achromatic bump peaking at $\sim T_0+3\times10^4$\,s \citep{2001ApJ...552L..35Z,2019ApJ...883...48L,2020ApJ...899..105L,2024ApJ...976L..20S}, however, due to the lack of an additional prompt emission, we think the scenario may be not proper for EP241217a.
Another model is an energetic jet edge component coming into view \citep{2019ApJ...873L...6S}, which can generate an achromatic bump and a jet break if the burst is observed nearly on-axis.
If EP241217a is observed on-axis, the prompt efficiency converting the total energy of the jet into $\gamma$-ray energy ($1-10^4$\,keV) is very low, $<1\%$, which is inferred to be $\sim10\%$ in the literature \citep{2016MNRAS.461...51B}.

\section{Conclusion}\label{sec:con}
The prompt emission of EP241217a was detected exclusively in the X-ray band; however, the lack of a $\gamma$-ray detection does not preclude the possibility that EP241217a is a Type II GRB.
As shown in the Amati relation diagram, the upper-limit curve for EP241217a coincides with the detection threshold for GRBs at redshifts $z > 4.5$.
Moreover, a significant number of GRBs with $E_{\gamma, {\rm iso}} \lesssim 10^{53}$,erg have been observed at lower redshifts, suggesting that EP241217a could be a high-redshift, intrinsically faint GRB whose $\gamma$-ray emission falls below the sensitivity of current $\gamma$-ray instruments.

The X-ray light curve reveals a plateau with $\alpha\approx0.4$ lasting for $\sim5\times10^4$\,s.
However, when combining optical data, we found the presence of a likely achromatic bump peaking at $T_0+3\times10^4$\,s.
Since the bump seems to be achromatic, the external origin of the bump is assumed, e.g., the rapid density enhancement of the circumburst medium.
This bump likely causes a significant overestimation of the plateau’s end time. 
Consequently, the overestimated end time places EP241217a as an outlier from the LTE relation observed in GRB X-ray afterglows by more than 2$\sigma$.

We modeled EP241217a under the scenario of a mildly relativistic jet coasting in a wind-like medium.
The best-fit parameters reveal the bulk Lorentz factor of the jet is $\Gamma\approx46$ and the jet is observed slightly off-axis.
However this scenario is not sufficient to produce a nice fit to the data (especially for the first $r'$-band detection, see Figure \ref{fig:modelLc}), and some issues do exist, like the model predicts a softer spectrum than observed during the plateau phase (please refer to Section \ref{sec:discussion} for details).
We further discussed alternative models explaining X-ray plateaus, but each encounters difficulties in reproducing key features of EP241217a.
Therefore, we conclude that according to our numerical result, the scenario of a mildly relativistic jet coasting in a wind-like medium is not sufficient to explain all observed features of EP241217a, and further theoretical modeling is encouraged to explore the origin of EP241217a.

\begin{acknowledgments}
This work is based on the data obtained with Einstein Probe, a space mission supported by Strategic Priority Program on Space Science of Chinese Academy of Sciences, in collaboration with ESA, MPE and CNES.
This work made use of data supplied by the UK Swift Science Data Centre at the University of Leicester.

H. Zhou thanks J. Ren, H. Wang and Y. Wang for useful discussion and modeling of EP241217a. 
Thanks to D. Xu, L. Xing, J. Rastinejad, A. J. Levan, P. K. Blanchard, W.-F. Fong, B. Gompertz, D. B. Malesani, C. D. Kilpatrick, G. P. Lamb, B. D. Metzger, M. Nicholl, N. Tanvir and the WFST collaboration for their contribution of optical data to this paper.
Thanks to C.-W. Wang for the contribution of GECAM-C data.
Thanks to J.-J. Geng for arranging WFST ToO observations and useful discussions. 
Thanks to B.-Y. Liu for the reduction of the WFST data.
H. Zhou, R.-D. Liang and S.-F. Zhu were transient advocates for EP241217a and Z.-X. Ling was the duty scientist for EP241217a.
Thanks to R. Yu, Z.-P. Jin and D.-M. Wei for suggestions and/or scientific oversights.

This work is supported by the Strategic Priority Research Program of the Chinese Academy of Sciences (grant No. XDB0550400), the National Key Research and Development Program of China (Nos. 2023YFA1608100, 2024YFA1611704), the National Natural Science Foundation of China (NSFC, Nos. 12225305, 12321003, 12473049), Jiangsu Provincial Double-Innovation Doctor Program (Grant No. JSSCBS0216), the Jiangsu Funding Program for Excellent Postdoctoral Talent (grant Nos. 2024ZB110, 2025ZB272), the Postdoctoral Fellowship Program (grant No. GZC20241916), the General Fund of the China Postdoctoral Science Foundation (grant No. 2024M763530, 2024M763531), and the Postdoctoral Innovation Talents Support Program (No. BX20250159, BX20250160).
\end{acknowledgments}

\bibliography{sample631}{}
\bibliographystyle{aasjournal}

\appendix
\section{Parameters of the best fitted model}\label{app:par}
Posterior distributions of best-fit parameters are shown in Figure \ref{fig:corner} with uncertainties are all at 1-$\sigma$ confidence level.

Fitted parameters are the density of the ISM-like medium $n_0$, the wind parameter $A_\star$, the isotropic kinetic energy of the jet $E_{k, {\rm iso}}$, the electron index $p$, the initial bulk Lorentz factor of the jet $\Gamma_0$, the micro parameter converting the internal energy of the jet to accelerated electrons $\epsilon_e$, the micro parameter converting the internal energy of the jet to magnetic fields $\epsilon_B$, the half jet opening angle $\theta_j$, the ratio of the viewing angle to the half jet opening angle $\theta_v/\theta_j$, the fraction of accelerated electrons to all electrons $\xi_e$, the color excess between B and V bands for the host galaxy $E(B-V)$, and the host absorption in the $r'$-band in magnitude $A_r$ (Appendix \ref{app:lyman}).
The posterior distribution is summarized in Table \ref{tbl:fit_par}.

The number density of the free-stellar-wind medium is given by $n_f = 3\times10^{35}A_{\star}r^{-2}\text{cm}^{-1}$.
The mass density of the free-to-shocked wind environment is described as
\begin{equation}
\label{eq:4}
n(r)=\left\{
\begin{array}{ll}
n_f (r), & r < r_{\rm{tr}}, \\
n_0 = \xi n_f (r_{\rm tr}), & r \geqslant r_{\rm tr},
\end{array}
\right.
\end{equation}
where $r_{\rm tr}$ is the transition radius of the circumburst environment, and $\xi=\frac{(\hat{\gamma}+1)\mathcal{M}_{\rm wind}^2}{(\hat{\gamma}-1)\mathcal{M}_{\rm wind}^2+2}=4$ is adopted in this paper, where we take $\hat{\gamma}=5/3$ for stellar wind and the Mach number $\mathcal{M}_{\rm wind} \gg 1$.

\section{The influence of the Lyman absorption}\label{app:lyman}
\begin{figure}
    \centering
    \includegraphics[width=0.49\linewidth]{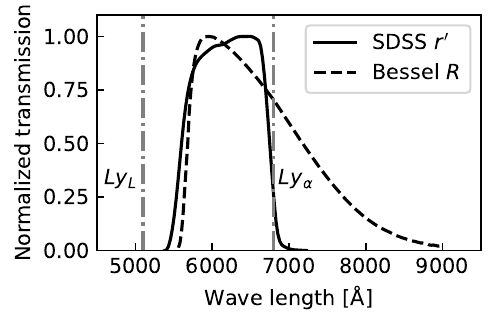}
    \caption{The normalized transmission curve of Sloan $r^\prime$ and Bessel $R$. The solid line and the dashed line represent the transmission curves of Sloan $r^\prime$ and Bessel $R$, respectively. The 2 vertical dot-dashed lines represent the Lyman limit $Ly_L=912(1+z)\,{\rm \AA}$ and the Lyman $\alpha$, $Ly_\alpha=1216(1+z)\,{\rm \AA}$.}
    \label{fig:filTrans}
\end{figure}

The redshift of EP241217a is $z=4.59$, and the central wavelength of the Lyman $\alpha$ ($Ly_\alpha$) is $1216(1+z)\tAA\approx6797\tAA$ in the observer frame.
Hence, the $r'/R$-band (with a central wavelength of $\sim6200/6500\tAA$) photometries may suffer from the Lyman absorption.
The transmission curve of SDSS $r'$ is adopted from \cite{2010AJ....139.1628D} and the transmission curve Bessel $R$ of NOT is downloaded from the official website\footnote{\url{https://www.not.iac.es/instruments/filters/filters.php}}.
Figure \ref{fig:filTrans} shows the normalized transmission curve, the $Ly_\alpha$, and the Lyman limit at the observer frame.
The SDSS $r'$ lies almost entirely in the Lyman forest, hence we adopted an additional fitting parameter $A_r$ to describe the absorption of $r'$-band data, which works as the extinction, i.e., $r'^{\rm,obs}=r'^{\rm,int}+A_r$, where $r'^{\rm,obs}$ is the observed $r'$-band photometry and $r'^{\rm,int}$ is the intrinsic photometry at the observer frame.
The red end of the Bessel $R$ extends up to $\sim9000\tAA$, hence photons with $7,000\tAA\lesssim\lambda\lesssim9,000\tAA$ are free of the $Ly_\alpha$ absorption.
As a result, the only $R$-band photometry ($\sim T_0+5\times10^4$\,s) is brighter than other $r'$-band photometries significantly.

\section{Emission features of a jet coasting in the wind-like medium}\label{app:jet}
Equation (\ref{equ:vch}-\ref{equ:gamma_c_app}) follows the \cite{1998ApJ...497L..17S}, and it is important to clarify three observers here.
\begin{itemize}
    \item Observer 1: an observer moving together with the jet, all physical parameters measured by this observer are noted with $'$, e.g., $t'$
    \item Observer 2: a still observer in the rest frame of the burst, and the observer is in the forward direction of the jet, parameters measured by this observer do not have any modifiers, e.g., $t$
    \item Observer 3: a distant observer (the cosmology is needed to be considered) in the forward direction of the jet, parameters measured by this observer are marked with a superscript $obs$, e.g., $t^{obs}$
\end{itemize}
The characteristic synchrotron frequency $\nu_{ch}^{obs}$ of an electron with a Lorentz factor of $\gamma'$ in a magnetic field $B'$ is
\begin{equation}\label{equ:vch}
    \nu_{ch}^{obs}\approx\frac{3}{4\pi}\gamma'^2\frac{e B'_\perp}{m_e c}\frac{\Gamma}{1+z^{obs}},
\end{equation}
where $e$ and $m_e$ are the charge and the mass of the electron, $\Gamma$ is bulk Lorentz factor of the jet, and $z^{obs}$ is the redshift of the burst.
The $B'_\perp$ is the magnetic field strength perpendicular to the velocity of the electron, $B'_\perp=B'{\rm sin}\,\alpha$, where $\alpha$ is the angle between the magnetic field and the velocity of the electron.
For simplify, the angular modification factor ${\rm sin}\,\alpha$ is neglected in following derivations.
The magnetic field strength is determined by\
\begin{equation}\label{equ:B}
    B'=\sqrt{32\pi\epsilon_B(\Gamma-1)\Gamma n m_p c^2},
\end{equation}
where $\epsilon_B$ is the fraction converting the shock internal energy to the magnetic field, $n$ is the density of the circumburst density, and $m_p$ is the mass of proton.
Assuming the distribution of accelerated electrons follows the powerlaw distribution, i.e., $N(\gamma')\propto\gamma'^{-p}$.
The minimal energy $\gamma_m'$ and the cooling energy $\gamma_c'$ of accelerated electrons are
\begin{equation}\label{equ:gamma_m}
    \gamma_m'=\frac{p-2}{p-1}\frac{\epsilon_e}{\xi_e}(\Gamma-1)\frac{m_p}{m_e},
\end{equation}
and
\begin{equation}\label{equ:gamma_c_exact}
    \gamma_c'=\frac{12\pi(1-\sqrt{1-\Gamma^{-2}})\Gamma(1+z^{obs})m_ec}{\sigma_{\rm T}t^{obs}B'^2},
\end{equation}
where $\epsilon_e$ is the fraction converting the shock internal energy to electrons, $\xi_e$ is the ratio of accelerated electrons to total electrons, $\sigma_{\rm T}=\frac{8\pi}{3}(\frac{e^2}{4\pi\epsilon_0 m_e c^2})^2$ is Thomson cross section of electrons, $\epsilon_0$ is the vacuum permittivity, and $t^{obs}$ is the elapsed time after the burst observed by Observer 3.
Please note that here we neglected the contribution of the Inverse Compton scattering, which accelerates the cooling of electrons, to the cooling energy.
Though the bulk Lorentz factor is a few of tens, the factor $(1-\sqrt{1-\Gamma^{-2}})$ in the cooling energy can be safely approximated with $1/2\Gamma^2$ (when $\Gamma<5$, the difference will be $>1\%$).
Hence, the cooling energy can be reduced to
\begin{equation}\label{equ:gamma_c_app}
    \gamma_c'\approx\frac{6\pi(1+z^{obs})m_ec}{\sigma_{\rm T}\Gamma t^{obs}B'^2}.
\end{equation}
The final piece to derive the scaling laws of the evolution of characteristic frequencies of synchrotron emissions is the density of the circumurst medium, which is assumed to be
\begin{equation}\label{equ:n}
    n=\frac{5\times10^{11}\,{\rm g\,cm^{-1}}}{m_p}A_\star r^{-2},
\end{equation}
where $A_\star$ is the unit-less wind parameter, and $r$ is the radius swept by the jet in centimeters.
If assuming the swept radius is much larger than the launched radius of the jet, the radius can be connected with the $t^{obs}$
\begin{equation}\label{equ:r}
    r=\frac{\sqrt{1-\Gamma^{-2}}}{1-\sqrt{1-\Gamma^{-2}}}\frac{t^{obs}}{1+z^{obs}}\approx\frac{2\Gamma^2 t^{obs}}{1+z^{obs}}.
\end{equation}

Since the jet is coasting, the bulk Lorentz factor is constant, i.e., $\Gamma\propto t^0$.
Hence, combing Equation (\ref{equ:B}-\ref{equ:r}) one can get
\begin{equation}
\begin{aligned}
    r&\propto t^{obs}, \\
    n&\propto {t^{obs}}^{-2}, \\
    B'&\propto {t^{obs}}^{-1}, \\
    \gamma_m'&\propto {t^{obs}}^0, \\
    \gamma_c'&\propto t^{obs}.
\end{aligned}
\end{equation}
Substitute the above equations into Equation (\ref{equ:vch}), temporal evolutions of the typical emission frequency of the synchrotron and the cooling frequency are
\begin{equation}
\begin{aligned}
    \nu_m^{obs}&\propto\gamma_m'^2 B'\propto {t^{obs}}^{-1}, \\
    \nu_c^{obs}&\propto\gamma_c'^2 B'\propto t^{obs}.\\
\end{aligned}
\end{equation}
As a result, in the scenario that the jet coasting in the wind-like medium, the $\nu_c^{obs}$ increases with $t^{obs}$ but the $\nu_m^{obs}$ decreases with ${t^{obs}}^{-1}$.

By adopting the notation $Q=10^xQ_x$ in the centimeter–gram–second system of units.
The numerical values are
\begin{equation}
\begin{aligned}
    \nu_m^{obs}&\approx(5.0\times10^{16}\,{\rm Hz})(\frac{p-2}{p-1})^2 \Gamma_{1.5}^2 A_\star^{1/2} \xi_e^{-2} \epsilon_{e,-1}^2 \epsilon_{B,-2}^{1/2} {t_3^{obs}}^{-1}, \\
    \nu_c^{obs}&\approx(3.5\times10^{12}\,{\rm Hz}) \Gamma_{1.5}^2 A_\star^{-3/2} \epsilon_{B,-2}^{-3/2} t_3^{obs}, \\
\end{aligned}
\end{equation}
which are consistent with the results obtained in \cite{2022NatCo..13.5611D}.
Hence, the typical values of $\nu_m^{obs}$ and $\nu_c^{obs}$ shows that for ${\rm a\ few}\times10^3\,{\rm s}$ after the burst, the synchrotron emission is in the fast cooling scenario, and the X-ray band could be higher than ${\rm max}(\nu_m^{obs},\ \nu_c^{obs})$.
Here we use same notations A-F adopted in \cite{2022NatCo..13.5611D} to mark different segmentation of the synchrotron emission spectrum.
The key features of the coasting phase are summarized for reference in Table \ref{tbl:scaling}.
For EP241217a, the typical emission frequency and the cooling frequency at $T_0+5\times10^3$\,s are
\begin{equation}
\begin{aligned}
    \nu_m^{obs}&\approx(1.8\times10^{15}\,{\rm Hz})(\frac{t^{obs}}{5,000\,{\rm s}})^{-1}, \\
    \nu_c^{obs}&\approx(6.6\times10^{13}\,{\rm Hz})(\frac{t^{obs}}{5,000\,{\rm s}}). \\
\end{aligned}
\end{equation}
As a result, from $T_0+5\times10^3\,{\rm s}$ to $T_0+5\times10^4\,{\rm s}$, the evolution of the X-ray band is from segmentation C to segmentation F, which give same scaling relation, i.e., $f_\nu^{obs}\propto {t^{obs}}^{-(p-2)/2}{\nu^{obs}}^{-p/2}\approx{t^{obs}}^{-0.4}{\nu^{obs}}^{-1.4}$.
While the observed spectral index during the period is $\beta_{\rm OX}\approx\beta_{\rm X}\approx0.9$, which is conflict with the model prediction.

In addition, for the X-ray band ($\sim10^{17}$\,Hz), the most common solution for the plateau phase are segmentation C, F and if the duration of the jet coasting in the wind-like medium is long enough, the X-ray emission will finally evolves into segmentation E, which implies a spectral hardening.
If the coasting phase ends in the segmentation C or F stage and the medium is still wind-like, the cooling frequency is still increasing with $\nu_c^{obs}\propto{t^{obs}}^{1/2}$ and the typical emission frequency is still decreasing with $\nu_m^{obs}\propto{t^{obs}}^{-3/2}$, hence the spectral hardening is also expected even after the plateau.
In this case, we adopted a mixture model of the wind and the ISM medium, and the ISM medium will dominate the evolution after the end of the plateau.
In the ISM medium, the cooling frequency and the typical emission decrease with $\nu_c^{obs}\propto{t^{obs}}^{-1/2}$ and $\nu_m^{obs}\propto{t^{obs}}^{-3/2}$, respectively.
If the X-ray plateau ends in the segmentation C or F and the jet enters the ISM-like medium, the spectral index after the plateau should be same as the the spectral index during the plateau phase.

As a result, we summarize that the jet coasting in the wind-like medium is indeed able to generate X-ray plateaus, and a spectral hardening after the plateau or a similar spectral index after the plateau is expected.
However, the spectrum of EP241217a became slight softer from $\beta_{\rm X}\sim0.9$ to $\sim1.1$, which is not expected by the jet coasting in the wind-like medium scenario.

\begin{table}[ht]
    \centering
    \begin{tabular}{ccccccc}
        \hline
        $T_{\rm start}-T_0$ & $T_{\rm stop}-T_0$ & EXP & INST & Model & $\Gamma_{\rm X}/kT~^a$ & cstat/dof \\
        (s) & (s) & (s) & & & (1/keV) & \\
        \hline
        \multirow{2}{*}{0.0} & \multirow{2}{*}{114.0} & \multirow{2}{*}{114.0} & \multirow{2}{*}{EP-WXT} & PL & $1.37^{+0.48}_{-0.45}$ & 20.32/20 \\
         & & & & BB & $0.40^{+0.09}_{-0.07}$ & 18.20/20 \\
        278.9 & 364.6 & 80.2 & EP-FXT & PL & $1.89\pm0.11$ & 33.58/46 \\
        364.6 & 584.2 & 219.5 & EP-FXT & PL & $1.73\pm0.10$ & 47.99/48 \\
        584.2 & 1,097.9 & 513.6 & EP-FXT & PL & $1.92\pm0.11$ & 44.80/41 \\
        4,912.9 & 7,242.6 & 2,220.9 & EP-FXT & PL & $1.94\pm0.09$ & 37.64/58 \\
        24,331.8 & 25,987.2 & 1,485.9 & \swift-XRT & PL & $2.14\pm0.41$ & 14.16/24 \\
        27,796.7 & 31,373.3 & 3,076.1 & EP-FXT & PL & $2.11\pm0.12$ & 35.69/40 \\
        \hline
    \end{tabular}
    \\
    a. The best fitted photon index $\Gamma$ of absorbed PL model or the best fitted temperature $kT$ of absorbed blackbody model in the unit of keV.
    \caption{The X-ray spectral analysis of EP241217a. The first three columns are start time of the time bin to construct spectra, the end time and the exposure time (on-source time). Column ``INST" and ``Model" represent what instrument the spectral data comes from and which model is applied to fit the data. The last 2 columns are the bested fitted parameters (the X-ray photon index $\Gamma_{\rm X}$ for PL and the temperature $kT$ for BB) for the fitting model and the \textit{cstat} per degree of freedom for fitting. For the WXT spectrum, it cannot claim the BB model is preferred than the PL model, since the BB models only gives a slightly better fit to the data.}
    \label{tbl:sed}
\end{table}

\begin{table}[h]
    \centering
    \begin{tabular}{ccc}
        \hline
        Parameter & Unit & Fitted \\
        \hline
        ${\rm lg}(E_{k, {\rm iso}})$ & erg & $55.22^{+0.48}_{-0.62}$ \\
        ${\rm lg}(\xi_e)$ & & $-0.36^{+0.23}_{-0.36}$ \\
        ${\rm lg}(\Gamma)$ & & $1.66^{+0.29}_{-0.17}$ \\
        $\theta_j$ & deg & $0.77^{+0.48}_{-0.57}$ \\
        $p$ & & $2.81^{+0.08}_{-0.10}$ \\ 
        ${\rm lg}(\epsilon_e)$ & & $-1.31^{+0.34}_{-0.42}$ \\
        ${\rm lg}(\epsilon_b)$ & & $-2.98^{+0.77}_{-1.13}$ \\
        \hline
        ${\rm lg}(n_0)$ & cm$^{-3}$ & $1.33^{+0.82}_{-1.06}$ \\
        ${\rm lg}(A_\star)$ & & $0.19^{+0.51}_{-0.52}$ \\
        $A_V$ & mag & $0.50^{+0.06}_{-0.18}$ \\
        $A_r$ & mag & $3.63^{+0.32}_{-0.80}$ \\
        $\theta_v$ & deg & $1.32^{+1.08}_{-1.22}$ \\
        \hline
    \end{tabular}
    \caption{The posterior distribution of afterglow modeling parameters. All uncertainties are reported at the 1-$\sigma$ confidence level. Meanings of parameters are presented in Appendix \ref{app:par}, excepting the extinction in the $V$-band of the host galaxy $A_V=R_V\times E(B-V)$ (the extinction curve of the SMC is adopted, thus $R_V=2.93$) and the viewing angle $\theta_v=(\theta_v/\theta_j)\times10^{{\rm lg}(\theta_j)}$. Parameters in the upper table describes the jet, and parameters in the lower table describe the environment surrounding the EP241217a and the geometry effect (the viewing angle).}
    \label{tbl:fit_par}
\end{table}

\begin{table}[ht]
    \centering
    \begin{tabular}{ccccccc}
        \hline
        $T_{\rm m}-T_0$ & Elapse & EXP & INS & Band & Mag & Ref \\
        (hr) & (s) & (s) & & & (AB) \\
        \hline
        1.13 & ... & 3$\times$180 & LCO/Sinistro & $r^\prime$ & 20.51$\pm$0.21 & GCN 38588$^(1)$ \\
        4.47 & 50 & 50 & Gemin-N/GMOS & $r^\prime$ & 21.42$\pm$0.06 & This Work \\
        8.00 & ... & 15$\times$300 & LOT/Sinistro & $r^\prime$ & 20.76$\pm$0.33 & GCN 38592 \\
        8.66 & ... & 300 & Mephisto & $r^\prime$ & 20.67$\pm$0.27 & GCN 38636 \\
        13.09 & ... & 5$\times$360 & GIT & $r^\prime$ & $>$20.32$^a$ & GCN 38612 \\
        13.91 & 1986 & 9$\times$180 & NOT/StanCAM & $R$ & 20.64$\pm$0.05 & This Work \\
        31.28 & ... & 3$\times$300 & Mephisto & $r^\prime$ & $>$22.17 & GCN 38636 \\
        44.84 & ... & 6$\times$200 & Liverpool/IO:O & $r^\prime$ & $>$21.52 & GCN 38633 \\
        9.57 & ... & 300 & Mephisto & $i^\prime$ & 18.96$\pm$0.06 & GCN 38636 \\
        11.10 & ... & 3$\times$360 & GIT & $i^\prime$ & 19.15$\pm$0.10 & GCN 38612 \\
        12.57 & ... & 5$\times$360 & GIT & $i^\prime$ & 19.54$\pm$0.11 & GCN 38612 \\
        15.62 & ... & 6$\times$100 & Liverpool/IO:O & $i^\prime$ & 19.88$\pm$0.09 & GCN 38615 \\
        31.57 & ... & 3$\times$300 & Mephisto & $i^\prime$ & 21.61$\pm$0.52 & GCN 38636 \\
        32.32 & 368 & 4$\times$60 & WFST & $i^\prime$ & 21.24$\pm$0.15 & This Work \\
        37.20 & 750 & 10$\times$60 & WFST & $i^\prime$ & 22.00$\pm$0.26 & This Work \\
        63.08 & ... & 6$\times$200 & Liverpool/IO:O & $i^\prime$ & $>$22.02 & GCN 38641 \\
        473.24 & 3450 & 9$\times$360 & NOT/ALFOSC & $i^\prime$ & $>$24.12 & This Work \\
        2.65 & 1088 & 6$\times$100 & Gemin-N/GMOS & $z^\prime$ & 19.67$\pm$0.06 & This Work \\
        8.58 & ... & 180 & Mephisto & $z^\prime$ & 18.45$\pm$0.13 & GCN 38636 \\
        12.05 & ... & 5$\times$360 & GIT & $z^\prime$ & $>$18.92$^a$ & GCN 38612 \\
        31.28 & ... & 3$\times$300 & Mephisto & $z^\prime$ & $>$20.34 & GCN 38636 \\
        \hline
    \end{tabular}
    \\
    $a$. 5-$\sigma$ up-limits. \\
    GCN 38588 \citep{2024GCN.38588....1I}, 38592 \citep{2024GCN.38592....1F}, 38612 \citep{2024GCN.38612....1M}, 38615 \citep{2024GCN.38615....1B}, 38633 \citep{2024GCN.38633....1K}, 38636 \citep{2024GCN.38636....1L}, 38641 \citep{2024GCN.38641....1B}, \\
    \caption{Optical photometries. Columns from left to right are the mid time of the observation since the $T_0$, the elapse time of the observation, the total exposure time, the instrument, the band and the AB magnitude. Upper limits are reported at 3-$\sigma$ level, otherwise stated. All detections and upper limits have been corrected for Milky Way extinction with $E(B-V)=0.1934$, i.e., $A_r=0.479$, $A_i=0.384$ and $A_z=0.285$.}
    \label{tbl:phot}
\end{table}

\begin{table}[h]
    \centering
    \begin{tabular}{ccc}
        \hline
        Notation & Seg. of the spectrum & Scaling relation \\
        \hline
        \multicolumn{3}{c}{Fast cooling} \\
        \hline
        A & $\nu<\nu_c<\nu_m$ & $f_\nu\propto {t}^{-1/3}{\nu}^{1/3}$ \\
        B & $\nu_c<\nu<\nu_m$ & $f_\nu\propto {t}^{1/2}{\nu}^{-1/2}$ \\
        C & $\nu_c<\nu_m<\nu$ & $f_\nu\propto {t}^{-(p-2)/2}{\nu}^{-p/2}$ \\
        \hline
        \multicolumn{3}{c}{Slow cooling} \\
        \hline
        D & $\nu<\nu_m<\nu_c$ & $f_\nu\propto {t}^{1/3}{\nu}^{1/3}$ \\
        E & $\nu_m<\nu<\nu_c$ & $f_\nu\propto {t}^{-(p-1)/2}{\nu}^{-(p-1)/2}$ \\
        F & $\nu_m<\nu_c<\nu$ & $f_\nu\propto {t}^{-(p-2)/2}{\nu}^{-p/2}$ \\
        \hline
    \end{tabular}
    \caption{The firs column represents notations to mark different segmentation of synchrotron spectrum (the second column). The last column presents the scaling relation. In this table, the superscript $^{obs}$ is not shown, i.e., all time, frequency and flux densities are measured in the observer frame (Observer 3 defined in Appendix \ref{app:jet}).}
    \label{tbl:scaling}
\end{table}

\end{document}